\documentclass[journal,a4paper,journal,onecolumn,draftclsnofoot,12pt]{IEEEtran}

\usepackage[english]{babel}
\usepackage[utf8]{inputenc}
\usepackage[T1]{fontenc}
\usepackage{hyphenat}

\usepackage{times}
\usepackage{amsmath,amssymb,amstext,amsfonts,mathrsfs,amsthm,mathtools,cases}
\usepackage{comment}

\usepackage{graphicx}                           
\graphicspath{{.}{./Figures/}}
\usepackage{subfig}

\usepackage[nolist,printonlyused]{acronym}      

\usepackage[sort]{cite}                         

\usepackage{enumerate}                          

\usepackage{multirow}
\usepackage{tabularx}
\usepackage{booktabs}
\newcolumntype{C}{>{\centering\arraybackslash}X}
\newcolumntype{R}{>{\flushright\arraybackslash}X}
\newcolumntype{L}{>{\flushleft\arraybackslash}X}

\usepackage{xspace} 
\newcommand*{\unit}[2]{\mbox{\ensuremath{#1\,\mathrm{#2}}}}

\usepackage{float}
\floatname{algorithm}{\footnotesize Algorithm}

\usepackage{url}

\theoremstyle{definition}

\usepackage{color}
\def\REVFR#1{\textcolor{black}{#1}}

\IEEEoverridecommandlockouts

\allowdisplaybreaks[4]

\begin{document}
\date{\today}

\title{Full-Duplex and Dynamic-TDD: \\
Pushing the Limits of Spectrum Reuse in Multi-Cell Communications}

\author{Jos\'{e} Mairton B.~da Silva Jr.$^{\star}$, Gustav Wikstr\"{o}m$^{\dagger}$, Ratheesh K. Mungara$^{\dagger}$, Carlo Fischione${^\star}$\\
$^\star$KTH Royal Institute of Technology, Stockholm, Sweden \\
$^\dagger$Ericsson Research, Kista, Sweden}


\begin{acronym}[LTE-Advanced]
  \acro{2G}{2$^\text{nd}$~generation}
  \acro{3-DAP}{3-Dimensional Assignment Problem}
  \acro{3G}{3$^\text{rd}$~generation}
  \acro{3GPP}{3$^\text{rd}$~Generation Partnership Project}
  \acro{4G}{4$^\text{th}$~generation}
  \acro{5G}{5$^\text{th}$~generation}
  \acro{AA}{Antenna Array}
  \acro{AC}{Admission Control}
  \acro{ACLR}{adjacent-channel leakage ratio}
  \acro{ACS}{adjacent-channel selectivity}
  \acro{AD}{Attack-Decay}
  \acro{ADC}{analog-to-digital converter}
  \acro{ADSL}{Asymmetric Digital Subscriber Line}
  \acro{AHW}{Alternate Hop-and-Wait}
  \acro{AMC}{Adaptive Modulation and Coding}
  \acro{AP}{Access Point}
  \acro{APA}{Adaptive Power Allocation}
  \acro{ARMA}{Autoregressive Moving Average}
  \acro{ATES}{Adaptive Throughput-based Efficiency-Satisfaction Trade-Off}
  \acro{AWGN}{additive white Gaussian noise}
  \acro{BB}{Branch and Bound}
  \acro{BCD}{block coordinate descent}
  \acro{BD}{Block Diagonalization}
  \acro{BER}{Bit Error Rate}
  \acro{BF}{Best Fit}
  \acro{BFD}{bidirectional full duplex}
  \acro{BLER}{BLock Error Rate}
  \acro{BPC}{Binary Power Control}
  \acro{BPSK}{Binary Phase-Shift Keying}
  \acro{BRA}{Balanced Random Allocation}
  \acro{BS}{base station}
  \acro{BSUM}{block successive upper-bound minimization}
  \acro{CAP}{Combinatorial Allocation Problem}
  \acro{CAPEX}{Capital Expenditure}
  \acro{CBF}{Coordinated Beamforming}
  \acro{CBR}{Constant Bit Rate}
  \acro{CBS}{Class Based Scheduling}
  \acro{CC}{Congestion Control}
  \acro{CDF}{cumulative distribution function}
  \acro{CDM}{code division multiplexing}
  \acro{CDMA}{Code-Division Multiple Access}
  \acro{CL}{Closed Loop}
  \acro{CLI}{cross-link interference}
  \acro{CLPC}{Closed Loop Power Control}
  \acro{CNR}{Channel-to-Noise Ratio}
  \acro{CNN}{convolutional neural network}
  \acro{CPA}{Cellular Protection Algorithm}
  \acro{CPICH}{Common Pilot Channel}
  \acro{CoMP}{Coordinated Multi-Point}
  \acro{CQI}{Channel Quality Indicator}
  \acro{CRM}{Constrained Rate Maximization}
	\acro{CRN}{Cognitive Radio Network}
  \acro{CS}{Coordinated Scheduling}
  \acro{CSI}{channel state information}
  \acro{CSMA}{carrier sense multiple access}
  \acro{CUE}{Cellular User Equipment}
  \acro{D2D}{device-to-device}
  \acro{DAC}{digital-to-analog converter}
  \acro{DCA}{Dynamic Channel Allocation}
  \acro{DE}{Differential Evolution}
  \acro{DFT}{Discrete Fourier Transform}
  \acro{DIST}{Distance}
  \acro{DL}{downlink}
  \acro{DMA}{Double Moving Average}
  \acro{DMRS}{demodulation reference signal}
  \acro{D2DM}{D2D Mode}
  \acro{DMS}{D2D Mode Selection}
  \acro{DPC}{Dirty Paper Coding}
  \acro{DRA}{Dynamic Resource Assignment}
  \acro{DSA}{Dynamic Spectrum Access}
  \acro{DSM}{Delay-based Satisfaction Maximization}
  \acro{D-TDD}{dynamic time-division duplexing}
  \acro{ECC}{Electronic Communications Committee}
  \acro{EFLC}{Error Feedback Based Load Control}
  \acro{EI}{Efficiency Indicator}
  \acro{eNB}{Evolved Node B}
  \acro{EPA}{Equal Power Allocation}
  \acro{EPC}{Evolved Packet Core}
  \acro{EPS}{Evolved Packet System}
  \acro{E-UTRAN}{Evolved Universal Terrestrial Radio Access Network}
  \acro{ES}{Exhaustive Search}
  \acro{FD}{full-duplex}
  \acro{FDD}{frequency-division duplex}
  \acro{FDM}{Frequency Division Multiplexing}
  \acro{FER}{Frame Erasure Rate}
  \acro{FF}{Fast Fading}
  \acro{FSB}{Fixed Switched Beamforming}
  \acro{FST}{Fixed SNR Target}
  \acro{FTP}{File Transfer Protocol}
  \acro{GA}{Genetic Algorithm}
  \acro{GBR}{Guaranteed Bit Rate}
  \acro{GLR}{Gain to Leakage Ratio}
  \acro{GOS}{Generated Orthogonal Sequence}
  \acro{GPL}{GNU General Public License}
  \acro{GRP}{Grouping}
  \acro{HARQ}{Hybrid Automatic Repeat Request}
  \acro{HD}{half-duplex}
  \acro{HMS}{Harmonic Mode Selection}
  \acro{HOL}{Head Of Line}
  \acro{HSDPA}{High-Speed Downlink Packet Access}
  \acro{HSPA}{High Speed Packet Access}
  \acro{HTTP}{HyperText Transfer Protocol}
  \acro{ICMP}{Internet Control Message Protocol}
  \acro{ICI}{Intercell Interference}
  \acro{ID}{Identification}
  \acro{IETF}{Internet Engineering Task Force}
  \acro{ILP}{Integer Linear Program}
  \acro{JRAPAP}{Joint RB Assignment and Power Allocation Problem}
  \acro{UID}{Unique Identification}
  \acro{IID}{Independent and Identically Distributed}
  \acro{IIR}{Infinite Impulse Response}
  \acro{ILP}{Integer Linear Problem}
  \acro{IMT}{International Mobile Telecommunications}
  \acro{INV}{Inverted Norm-based Grouping}
	\acro{IoT}{Internet of Things}
  \acro{IP}{Integer Programming}
  \acro{IPv6}{Internet Protocol Version 6}
  \acro{ISD}{Inter-Site Distance}
  \acro{ISI}{Inter Symbol Interference}
  \acro{ITU}{International Telecommunication Union}
  \acro{JAFM}{joint assignment and fairness maximization}
  \acro{JAFMA}{joint assignment and fairness maximization algorithm}
  \acro{JOAS}{Joint Opportunistic Assignment and Scheduling}
  \acro{JOS}{Joint Opportunistic Scheduling}
  \acro{JP}{Joint Processing}
	\acro{JS}{Jump-Stay}
  \acro{KKT}{Karush-Kuhn-Tucker}
  \acro{L3}{Layer-3}
  \acro{LAC}{Link Admission Control}
  \acro{LA}{Link Adaptation}
  \acro{LC}{Load Control}
  \acro{LOS}{line of sight}
  \acro{LP}{Linear Programming}
  \acro{LTE}{Long Term Evolution}
	\acro{LTE-A}{\ac{LTE}-Advanced}
  \acro{LTE-Advanced}{Long Term Evolution Advanced}
  \acro{M2M}{Machine-to-Machine}
  \acro{MAC}{medium access control}
  \acro{MANET}{Mobile Ad hoc Network}
  \acro{MC}{Modular Clock}
  \acro{MCS}{Modulation and Coding Scheme}
  \acro{MDB}{Measured Delay Based}
  \acro{MDI}{Minimum D2D Interference}
  \acro{MF}{Matched Filter}
  \acro{MG}{Maximum Gain}
  \acro{MH}{Multi-Hop}
  \acro{MIMO}{multiple-input multiple-output}
  \acro{MINLP}{mixed integer nonlinear programming}
  \acro{MIP}{Mixed Integer Programming}
  \acro{MISO}{Multiple Input Single Output}
  \acro{MLWDF}{Modified Largest Weighted Delay First}
  \acro{MME}{Mobility Management Entity}
  \acro{MMSE}{minimum mean squared error}
  \acro{mmWave}{millimeter wave}
  \acro{MOS}{Mean Opinion Score}
  \acro{MPF}{Multicarrier Proportional Fair}
  \acro{MRA}{Maximum Rate Allocation}
  \acro{MR}{Maximum Rate}
  \acro{MRC}{Maximum Ratio Combining}
  \acro{MRT}{maximum ratio transmission}
  \acro{MRUS}{Maximum Rate with User Satisfaction}
  \acro{MS}{Mode Selection}
  \acro{MSE}{mean squared error}
  \acro{MSI}{Multi-Stream Interference}
  \acro{MTC}{Machine-Type Communication}
  \acro{MTSI}{Multimedia Telephony Services over IMS}
  \acro{MTSM}{Modified Throughput-based Satisfaction Maximization}
  \acro{MU-MIMO}{Multi-User Multiple Input Multiple Output}
  \acro{MU}{Multi-User}
  \acro{NAS}{Non-Access Stratum}
  \acro{NB}{Node B}
  \acro{NCL}{Neighbor Cell List}
  \acro{NLP}{Nonlinear Programming}
  \acro{NLOS}{non-line of sight}
  \acro{NMSE}{Normalized Mean Square Error}
  \acro{NORM}{Normalized Projection-based Grouping}
  \acro{NP}{non-polynomial time}
  \acro{NR}{New Radio}
  \acro{NRT}{Non-Real Time}
  \acro{NSPS}{National Security and Public Safety Services}
  \acro{O2I}{Outdoor to Indoor}
  \acro{OFDMA}{Orthogonal Frequency Division Multiple Access}
  \acro{OFDM}{orthogonal frequency division multiplexing}
  \acro{OFPC}{Open Loop with Fractional Path Loss Compensation}
	\acro{O2I}{Outdoor-to-Indoor}
  \acro{OL}{Open Loop}
  \acro{OLPC}{Open-Loop Power Control}
  \acro{OL-PC}{Open-Loop Power Control}
  \acro{OPEX}{Operational Expenditure}
  \acro{ORB}{Orthogonal Random Beamforming}
  \acro{JO-PF}{Joint Opportunistic Proportional Fair}
  \acro{OSI}{Open Systems Interconnection}
  \acro{PAIR}{D2D Pair Gain-based Grouping}
  \acro{PAPR}{Peak-to-Average Power Ratio}
  \acro{P2P}{Peer-to-Peer}
  \acro{PC}{Power Control}
  \acro{PCI}{Physical Cell ID}
  \acro{PDCCH}{physical downlink control channel}
  \acro{PDD}{penalty dual decomposition}
  \acro{PDF}{Probability Density Function}
  \acro{PER}{Packet Error Rate}
  \acro{PF}{Proportional Fair}
  \acro{P-GW}{Packet Data Network Gateway}
  \acro{PHY}{physical}
  \acro{PL}{Pathloss}
  \acro{PRB}{Physical Resource Block}
  \acro{PROJ}{Projection-based Grouping}
  \acro{ProSe}{Proximity Services}
  \acro{PS}{phase shifter}
  \acro{PSO}{Particle Swarm Optimization}
  \acro{PUCCH}{physical uplink control channel}
  \acro{PZF}{Projected Zero-Forcing}
  \acro{QAM}{quadrature amplitude modulation}
  \acro{QoS}{quality of service}
  \acro{QPSK}{Quadri-Phase Shift Keying}
  \acro{RAISES}{Reallocation-based Assignment for Improved Spectral Efficiency and Satisfaction}
  \acro{RAN}{Radio Access Network}
  \acro{RA}{Resource Allocation}
  \acro{RAT}{Radio Access Technology}
  \acro{RATE}{Rate-based}
  \acro{RB}{resource block}
  \acro{RBG}{Resource Block Group}
  \acro{REF}{Reference Grouping}
  \acro{RF}{radio-frequency}
  \acro{RLC}{Radio Link Control}
  \acro{RM}{Rate Maximization}
  \acro{RNC}{Radio Network Controller}
  \acro{RND}{Random Grouping}
  \acro{RRA}{Radio Resource Allocation}
  \acro{RRM}{Radio Resource Management}
  \acro{RSCP}{Received Signal Code Power}
  \acro{RSRP}{reference signal receive power}
  \acro{RSRQ}{Reference Signal Receive Quality}
  \acro{RR}{Round Robin}
  \acro{RRC}{Radio Resource Control}
  \acro{RSSI}{received signal strength indicator}
  \acro{RT}{Real Time}
  \acro{RU}{Resource Unit}
  \acro{RUNE}{RUdimentary Network Emulator}
  \acro{RV}{Random Variable}
  \acro{SAC}{Session Admission Control}
  \acro{SCM}{Spatial Channel Model}
  \acro{SC-FDMA}{Single Carrier - Frequency Division Multiple Access}
  \acro{SD}{Soft Dropping}
  \acro{S-D}{Source-Destination}
  \acro{SDPC}{Soft Dropping Power Control}
  \acro{SDMA}{Space-Division Multiple Access}
  \acro{SDR}{semidefinite relaxation}
  \acro{SDP}{semidefinite programming}
  \acro{SER}{Symbol Error Rate}
  \acro{SES}{Simple Exponential Smoothing}
  \acro{S-GW}{Serving Gateway}
  \acro{SGD}{stochastic gradient descent}
  \acro{SINR}{signal-to-interference-plus-noise ratio}
  \acro{SI}{self-interference}
  \acro{SIP}{Session Initiation Protocol}
  \acro{SISO}{Single Input Single Output}
  \acro{SIMO}{Single Input Multiple Output}
  \acro{SIR}{Signal to Interference Ratio}
  \acro{SLNR}{Signal-to-Leakage-plus-Noise Ratio}
  \acro{SMA}{Simple Moving Average}
  \acro{SNR}{signal-to-noise ratio}
  \acro{SORA}{Satisfaction Oriented Resource Allocation}
  \acro{SORA-NRT}{Satisfaction-Oriented Resource Allocation for Non-Real Time Services}
  \acro{SORA-RT}{Satisfaction-Oriented Resource Allocation for Real Time Services}
  \acro{SPF}{Single-Carrier Proportional Fair}
  \acro{SRA}{Sequential Removal Algorithm}
  \acro{SRS}{sounding reference signal}
  \acro{SU-MIMO}{Single-User Multiple Input Multiple Output}
  \acro{SU}{Single-User}
  \acro{SVD}{Singular Value Decomposition}
  \acro{SVM}{support vector machine}
  \acro{TCP}{Transmission Control Protocol}
  \acro{TDD}{time-division duplex}
  \acro{TDMA}{Time Division Multiple Access}
  \acro{TNFD}{three node full duplex}
  \acro{TETRA}{Terrestrial Trunked Radio}
  \acro{TP}{Transmit Power}
  \acro{TPC}{Transmit Power Control}
  \acro{TTI}{transmission time interval}
  \acro{TTR}{Time-To-Rendezvous}
  \acro{TSM}{Throughput-based Satisfaction Maximization}
  \acro{TU}{Typical Urban}
  \acro{UE}{user equipment}
  \acro{UEPS}{Urgency and Efficiency-based Packet Scheduling}
  \acro{UL}{uplink}
  \acro{UMTS}{Universal Mobile Telecommunications System}
  \acro{URI}{Uniform Resource Identifier}
  \acro{URM}{Unconstrained Rate Maximization}
  \acro{VR}{Virtual Resource}
  \acro{VoIP}{Voice over IP}
  \acro{WAN}{Wireless Access Network}
  \acro{WCDMA}{Wideband Code Division Multiple Access}
  \acro{WF}{Water-filling}
  \acro{WiMAX}{Worldwide Interoperability for Microwave Access}
  \acro{WINNER}{Wireless World Initiative New Radio}
  \acro{WLAN}{Wireless Local Area Network}
  \acro{WMMSE}{weighted minimum mean square error}
  \acro{WMPF}{Weighted Multicarrier Proportional Fair}
  \acro{WPF}{Weighted Proportional Fair}
  \acro{WSN}{Wireless Sensor Network}
  \acro{WWW}{World Wide Web}
  \acro{XIXO}{(Single or Multiple) Input (Single or Multiple) Output}
  \acro{ZF}{zero forcing}
  \acro{ZMCSCG}{Zero Mean Circularly Symmetric Complex Gaussian}
\end{acronym}

\maketitle
\IEEEpeerreviewmaketitle

\begin{abstract}
Although in cellular networks full-duplex and dynamic time-division duplexing promise increased spectrum efficiency, their potential is so far challenged by increased interference.
While previous studies have shown that self-interference can be suppressed to a sufficient level,
we show that the cross-link interference for both duplexing modes, especially from base station to base station, is the remaining challenge in multi-cell networks, restricting the uplink performance. 
Using beamforming techniques of low-complexity, we show that this interference can be mitigated, and that full-duplex and dynamic time-division duplexing can substantially increase the capacity of multi-cell networks. 
Our results suggest that if we can control the cross link interference in full-duplex, then we can almost double the multi cell network capacity as well as user throughput. 
Therefore, the techniques in this paper have the potentiality to enable a smooth introduction of full-duplex into cellular systems. 
\end{abstract}

\section{Introduction}\label{sec:intro}
Cellular networks have evolved from utilizing higher frequency reuse factors to universal frequency reuse, i.e., neighboring cells utilize the same frequency band in the same link direction, separated in frequency or in time. 
To meet the service requirements due to the ever-increasing mobile data traffic~\cite{EricssonReport2018}, 
it is time to rethink the current reuse-1 and push the limits of spectrum reuse further.
Although \ac{MIMO} transmission is already used to enhance the spectral efficiency by means of frequency reuse across spatial or antenna dimensions, with the extension of multi-user MIMO (MU-MIMO)\acused{MU-MIMO}, we focus on a parallel approach.
Specifically, we consider \ac{D-TDD} transmissions which increase the resource efficiency, and \ac{FD} transmissions which have the potential to double the system capacity by means of reuse-1/2.

D-TDD is similar to static-\ac{TDD} in using one direction at a time in a cell, but in D-TDD this direction can be chosen dynamically, and in theory exactly adapted to the traffic needs. This straightforward extension of static-\ac{TDD} would not only yield 100\% capacity improvement, but would also increase efficiency because all resources can be given to a user when it is scheduled~\cite{Shen2012}. 
In contrast, \ac{FD} systems in which the \ac{BS} is \ac{FD} and the users are \ac{HD}, are reuse-1/2 because they can use the same time-frequency twice in a cell to schedule users in both \ac{UL} and \ac{DL} directions. 
Realizing such FD networks in practice would be extremely useful to increase capacity in heavily loaded networks~\cite{Sabharwal2014}.

In current \ac{FDD} or static-\ac{TDD} configurations, the interference is of the same type as the signal: from other \acp{BS} in the \ac{DL} or from other \acp{UE} in the \ac{UL}. 
This normally ensures that the signal level is relatively higher than each interference level, and links can be maintained through adequate modulation and coding. 
However, since only one direction is set for a certain time-frequency resource, the spectral efficiency is limited to that of reuse-1 systems. 
At most, the share of \ac{UL} or \ac{DL} resources can only be partially allocated to a user, but not all.
While the spectrum usage is further optimized by means of \ac{D-TDD} and \ac{FD} communications, additional interference management is necessary in order to harvest these benefits.

\REVFR{In this paper, we analyze the importance of the inter-cell interference suppression for \ac{FD} and \ac{D-TDD} transmissions.
Specifically, we show that the interference between \acp{BS} is the limiting factor of the performance of \ac{FD} and \ac{D-TDD}.}
We analyze current interference mitigation techniques from both academia and industry, and propose a low-complexity solution to mitigate the interference from neighboring \acp{BS}.
The numerical results show high throughput gains for the system and individual \ac{UL} and \ac{DL} users on different traffic scenarios.
With proper inter-cell interference mitigation, we show that \ac{FD} transmissions are close to the theoretical doubling of the system throughput.

In Section~\ref{sec:fd_dtdd}, the interference situation of \ac{D-TDD} and \ac{FD} is analyzed and the \ac{CLI} between \ac{BS}s is identified as the limiting factor. In Section~\ref{sec:mimo_multi_cell}, mitigation techniques for the interference are discussed, and in Section~\ref{sec:num_res} it is shown how null-forming in the \ac{BS} transmissions can suppress the interference and bring \ac{D-TDD} and \ac{FD} into an efficient region. 
Finally, Section~\ref{sec:concl} gives an overview of the lessons learned and perspectives for future works.

\section{Overview of Dynamic-TDD and Full-Duplex Communications}\label{sec:fd_dtdd}

\subsection{Interference types}\label{sub:interference}

\begin{figure}
	\centering
	\includegraphics[width=0.7\linewidth,trim=5mm 5mm 5mm 11mm,clip]{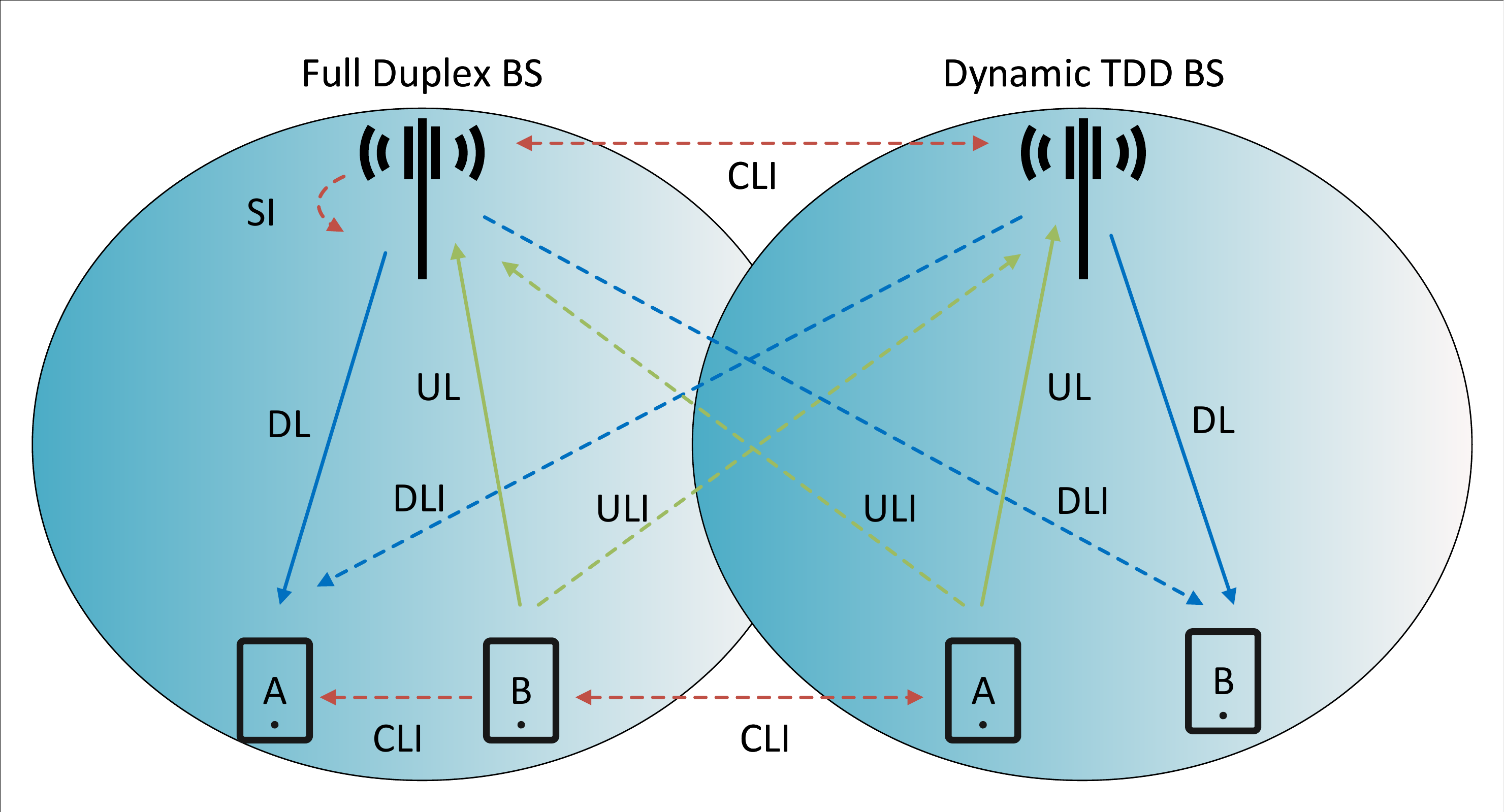}
	\caption{Interference scenarios in \ac{FD} (left) and \ac{D-TDD} (right) multi-cell environments. Note that in \ac{D-TDD}, \REVFR{\acp{UE}} A and B are not active at the same time; whereas in \ac{FD} both \acp{UE} are active at the same time. Moreover, DLI and ULI represent \ac{DL} and \ac{UL} interference, respectively.
	}
	\label{fig:FD_dynamic_TDD}
\end{figure}

\ac{D-TDD} cellular systems experience two kinds of inter-cell interference beyond those of static-\ac{TDD}: \ac{BS}-to-\ac{BS} interference between \ac{DL} and \ac{UL} \acp{BS}, and the \ac{UE}-to-\ac{UE} interference between users in \ac{UL} and \ac{DL} cells~\cite{Shen2012}. 
Both types of interference are referred to as inter-cell \ac{CLI} (cf. Figure \ref{fig:FD_dynamic_TDD}).
In \ac{FD} cellular systems with a \ac{FD} \ac{BS} and \ac{HD} users, the additional interference beyond those of \ac{D-TDD} are the \ac{SI}, i.e., intra-cell interference from the BS to itself, and the intra-cell \ac{UE}-to-\ac{UE} interference from \ac{UL} users to \ac{DL} users, referred to as intra-cell \ac{CLI}.
To harvest the benefits of \ac{D-TDD} and \ac{FD}, these additional interference sources need to be mitigated.
Inter-cell \ac{CLI} is present in both duplexes, meaning that a solution for managing \ac{CLI} in \ac{D-TDD} would help for \ac{FD}, and vice versa. As we will show, setting \ac{SI} aside, inter-cell \ac{CLI} is the dominant issue, and once an efficient suppression is in place, good performance can be achieved for \ac{FD} and \ac{D-TDD}.

\subsection{Dynamic-TDD}\label{sub:dynamic_tdd}

Unlike \ac{FDD}, \ac{TDD} cellular systems have the potential to increase efficiency by adapting the \ac{UL} and \ac{DL} time resources according to the traffic load. 
By using different TDD patterns dynamically across cells, under lower-traffic loads, near doubling of bit-rates over \ac{FDD} can be expected under lower-traffic loads owing to the wider bandwidth used.
At higher loads, queuing delay for the desired direction is expected to reduce the performance down to that of \ac{FDD}, and more so if traffic is highly asymmetric in one direction.
As shown in the right part of Figure~\ref{fig:FD_dynamic_TDD}, \ac{D-TDD} allows for a scheduler to dynamically set the direction to serve UL user-A or DL user-B.

The authors in~\cite{FD-smallcell} characterized the impact of \ac{CLI} in terms of the statistics of \ac{SINR} and spectral efficiency for a small-cell network. Specifically, by transforming the part of strong \ac{DL} interference into weak \ac{UL} interference, \ac{D-TDD} is shown to improve the \ac{DL} \ac{SINR} with respect to static-\ac{TDD}. Conversely, 
the \ac{UL} performance of \ac{D-TDD} is severely impacted as part of the weak \ac{UL} interference becomes strong \ac{DL} interference. 

\subsection{Full-Duplex Communications}\label{sub:full_duplex}

\ac{FD} has not been considered a viable communication solution so far due to high \ac{SI}. Recent advancements in antenna and analog/digital \ac{SI} cancellation techniques demonstrate suppression up to \unit{110}{dB}~\cite{Zhang2016, Everett2016}.
To achieve the required \ac{SI} suppression with reasonable form-factors,
\ac{FD} is primarily studied in the context of short-range communication, featuring lower transmit powers such as in WiFi/small-cells, by utilizing multiple antennas. 
It is important to understand the challenges involved in adapting full-duplex from an isolated link to the network-level.

A viable system-level solution---in terms of hardware complexity and form-factors at the transceivers---is to use full-duplex only at the multi-antenna \acp{BS}, while the \ac{UE}s are still operating in \ac{HD} mode.
Left part of Figure~\ref{fig:FD_dynamic_TDD} shows this setup, with one user-B in the \ac{UL} and one user-A in the \ac{DL}. 
Using \ac{FD} communication in the \ac{BS}, there is both intra- and inter-cell \ac{CLI} on \ac{UL}-\ac{DL} users, \ac{SI} at each \ac{BS}, and the \ac{BS}-to\ac{BS}. 
In \ac{D-TDD}, there is a trade-off between the \ac{DL}-to-\ac{UL} and \ac{UL}-to-\ac{DL} interference, i.e., an increase in one type causes a decrease in the other.
However, this is not the case in \ac{FD} networks because the number of interference sources increases.
Therefore, the \ac{SI}, \ac{CLI}, and the conventional intra- and inter-cell interference make the system-level operation of \ac{FD} very challenging.

\subsection{Design Questions and Challenges}\label{sub:design_chall}

\begin{figure}
	\centering
	\includegraphics[width=0.7\linewidth,trim=0mm 0mm 0mm 0mm,clip]{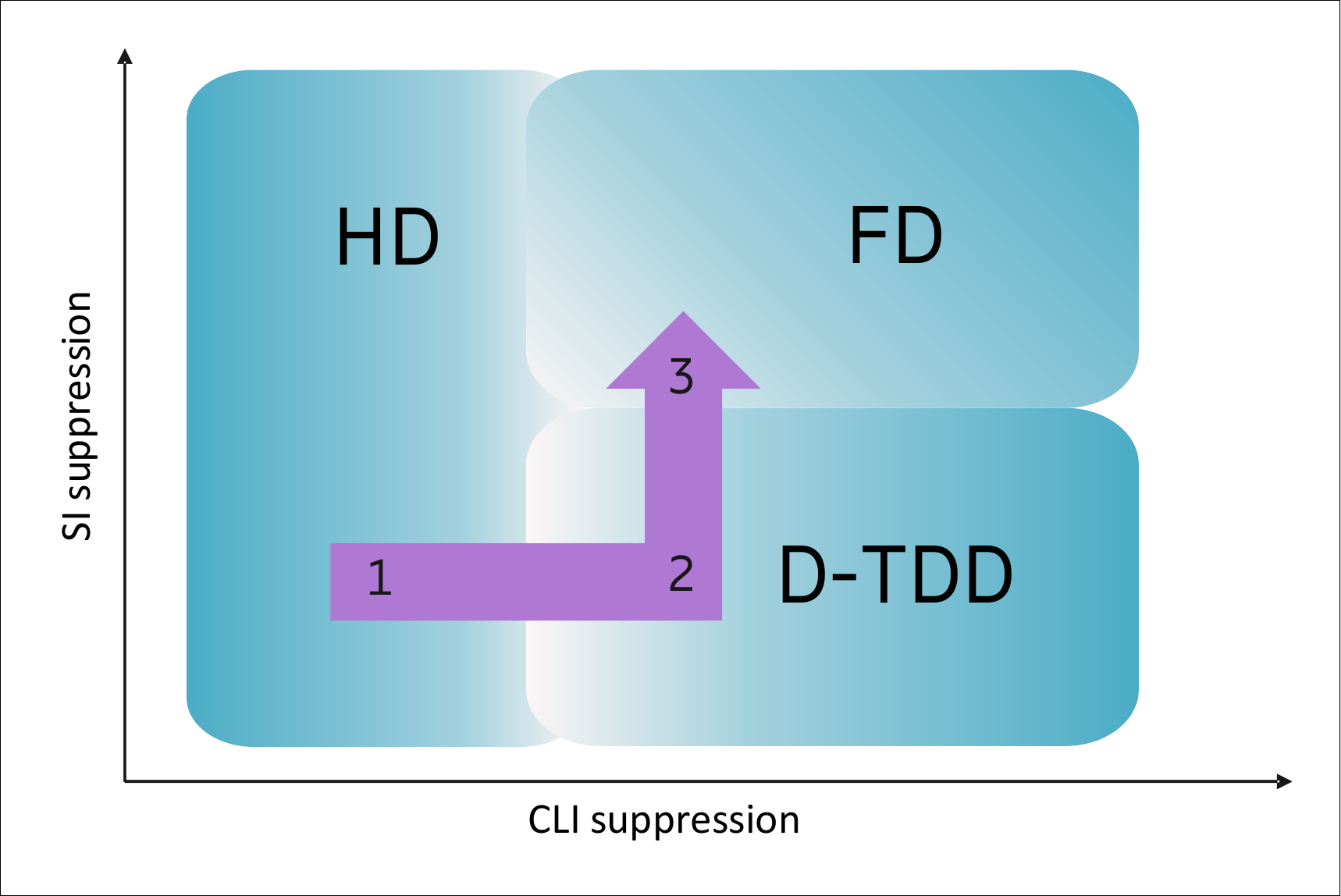}
	\caption{\REVFR{Regions of expected best use of \ac{HD}, \ac{D-TDD}, and \ac{FD} in multi-cell networks as function 
	of suppression capability}. Suppressing \ac{CLI} enables \ac{D-TDD}, while \ac{FD} requires suppression of the 
	\ac{SI} as well. The arrow indicates steps of expected technical advancement.}
	\label{fig:FD_DTDD_region}
\end{figure}

\ac{FD} communications experience more interference sources than \ac{D-TDD} in multi-cell scenarios. 
Nevertheless, due to the low-power transmission between \ac{UE}s and the \ac{SI} cancellation advancements, the \ac{CLI}, especially the \ac{BS}-to-\ac{BS} interference, can be the key challenge on the application of \ac{FD} communications in current cellular systems. Figure~\ref{fig:FD_DTDD_region} shows, along the arrows, the expected best use of \ac{HD}, \ac{D-TDD}, and \ac{FD} with respect to the \ac{SI} and \ac{CLI} suppression \REVFR{as function of suppression capability}.
The arrow starts at \ac{HD} technologies, either static-\ac{TDD} or \ac{FDD}, because in this case it is not necessary to suppress either \ac{SI} or the \ac{CLI}.
Notice through the x-axis that both duplexing technologies need to address \ac{CLI}, specifically the isolation between \acp{BS}.
Meanwhile in the y-axis, \ac{FD} also needs to address the further challenge of \ac{SI}, but only suppressing \ac{SI} does not suffice to enable use of \ac{FD}.
Since both technologies share a similar challenge, the path towards the implementation of a tighter spectrum reuse with \ac{FD} in cellular systems goes through the solution of \ac{BS} isolation proposed for the \ac{D-TDD} technology, which is represented by the arrow.

The strong CLI can be mitigated by various means such as coordinated beamforming and scheduling, power control, and hybrid dynamic/static \ac{DL}/\ac{UL} resource assignment. 
However, if two \ac{D-TDD} or \ac{FD} networks operate on adjacent-channels, additional interference appear that can only be handled by suppression of adjacent-channel and interference and receiver selectivity, e.g., \ac{ACLR} and \ac{ACS} \cite{3gpp.38.104}.
Therefore, to deploy either \ac{D-TDD} or \ac{FD} technologies, it is essential to investigate RF coexistence mechanisms among different operators in co-channel and adjacent-channels. 
Furthermore, the design of advanced mechanisms enabling interference measurements across \ac{BS}-to-\ac{BS} links and \ac{UL}-to-\ac{DL} links should be considered, which could enable efficient coexistence and handle \ac{CLI}.

Alternatively, \ac{CLI} can be managed with \ac{CSMA}, which is how \ac{D-TDD} is enabled in WiFi, but this in practice increases the reuse since nearby cells are prevented from transmitting simultaneously.

\section{Recent Advancements Relevant for FD and D-TDD}\label{sec:mimo_multi_cell}

\subsection{Multi-Cell Interference Cancellation Techniques}

Given the growing prevalence of \ac{MIMO} technology, it is worth to understand different interference cancellation techniques for \ac{D-TDD} and \ac{FD} that exploit spatial pre- and post-processing of the transmit and received signals.

The work in \cite{reaping-dynamicTDD} considers massive \ac{MIMO} at the BSs featuring large number of antennas and shows that the BS-to-BS interference can be made arbitrarily small and bounded in principle. On the other hand, \cite{3Dbeamforming-DTDD} exploits 3D beamforming with antenna horizontal and vertical radiation patterns being dependant on the spatial distribution of users’ locations. Specifically, the DL-to-UL and the UL-to-DL interference-to-signal ratios are characterized based on the analytical model developed for 3D beamforming.

In~\cite{IA-PC-FD}, the authors exploit a combination of MIMO interference alignment and power control to support self-backhauling in small-cell networks.
Linear programming is utilized to derive the feasibility conditions for interference alignment in a full-duplex network 
while power control is solved via convex optimization with the goal of maximizing the sum rate.
In~\cite{Koh2018}, the authors consider linear beamforming techniques for massive MIMO systems, and analyze the pilot overhead problem for cooperative and non-cooperative multi-cell scenarios using a central unit through the fronthaul. The results indicate that \ac{FD} outperforms \ac{HD} with an increasing number of transceiver \acl{RF} chains, and that the fronthaul capacity limits the sum rate and accuracy of the channel estimation in the cooperative \ac{FD} system.
\REVFR{In~\cite{Choi2013}, the authors propose an interference cancellation for the \ac{BS}-to-\ac{BS} interference based on null forming the elevation angle at the \ac{BS} antennas.
Due to broad cancellation over the azimuth and into a specific and predetermined elevation angle, there are losses in the signal power in the direction of elevation angle.}
In~\cite{Al-Saadeh2017}, the authors compare the performance of \ac{HD}, \ac{D-TDD}, and \ac{FD} for an indoor scenario  operating on millimeter-wave and using successive interference cancellation to remove the strongest interference from the \ac{BS} and users. The results indicate that \ac{FD} outperforms \ac{D-TDD} and \ac{HD} depending on the \ac{SI} cancellation level, and when the access point power is transmitting on a similar level as the \ac{UL} users. 

In contrast to the above complicated schemes, it is desirable to analyze the practical solutions proposed in the context of \ac{D-TDD} transmissions and tailor them to \ac{FD} communications, which is precisely discussed in the next section.

\subsection{3GPP 5G NR Studies on RF coexistence in \ac{D-TDD} Networks}\label{sub:3GPP_DTDD_studies}

\ac{5G} \ac{NR} aimed to support dynamic \ac{DL}/\ac{UL} assignments based on the instantaneous traffic demands. However, none of the \ac{CLI} management techniques and coexistence requirements are explicitly included in the initial release-15 specifications. To address this, the \ac{CLI} management including the coexistence mechanisms among different operators in adjacent cells has been considered within the scope of release-16. Specifically, subject to the minimum requirements on the levels of ACLR and ACS at the BS and at the UE, the performance (\ac{SINR} and throughput) degradation is characterized in various combinations of adjacent networks formed over macro and indoor scenarios. The following recommendations are provided based on the simulations conducted by various companies~\cite{3gpp.38.828}: 
\begin{itemize}
\item \ac{D-TDD} does not seem to be viable due to the performance degradation observed in the macro-to-macro scenario;
\item Two \ac{D-TDD} networks with a sufficient isolation operating respectively over indoor and macro scenarios do not cause performance degradation;
\item Similarly, two \ac{D-TDD} indoor networks do not cause any performance degradation, provided the BS and UE power levels are on the same order and the BS are placed judiciously by means of coordination between operators.
\end{itemize}
Therefore, \ac{D-TDD} with the above-mentioned measures can be used in indoor scenarios, but not in macro scenarios.

\section{Towards Full-Duplex Communications through Dynamic-TDD}\label{sec:num_res}

\subsection{System performance in \ac{UL} with and without interference suppression}\label{sub:UMa_results}

To gain insights about the required interference suppression and traffic dependence, we conduct system-level performance comparison of \ac{HD}, \ac{D-TDD}, and \ac{FD} networks, with focus on relative comparison rather than absolute numbers. 
We consider a simulation of a macro-cell network consisting of 7 tri-sector \acp{BS}, placed $500$ m apart in a hexagonal-grid. Each BS is equipped with an $8 \times 8$ cross-polarized antenna array, using $2 \times 2$ single user MIMO. The BS antenna height is $25$ m and transmit power is $40$ W, while the carrier frequency is $3.5$ GHz and system bandwidth is $40$ MHz. 
Notice that in \ac{FD} communications, $40$ MHz is reused for both \ac{UL} and \ac{DL} and split separately between the \ac{UL} and \ac{DL} users.
For \ac{D-TDD}, $40$ MHz also is used for \ac{UL} and \ac{DL} and split jointly among all \ac{UL} and \ac{DL} users in the cell.
Conversely, the \ac{HD} mode considers $20$ MHz \ac{FDD} band separately for \ac{UL} and \ac{DL}, which is divided between \ac{UL} and \ac{DL} users in the cell.
To gather statistics, $3000$ \ac{UE}s are deployed randomly outdoors with an antenna height of \unit{1.5}{m} \REVFR{and maximum \unit{0.2}{W} power targeting \unit{10}{dB} \ac{SNR} at the \ac{BS}}. The pathloss follows the ITU urban-macro outdoor scenario~\cite{ITURM2135}, corresponding to good coverage yet highly unfavorable to \ac{FD} communication.
We assume the equal traffic load across the users and that the \ac{DL} load equals the twice of the \ac{UL} load in an uncorrelated manner and each user experiences a set of random interferers subject to the traffic levels in the \ac{DL} and \ac{UL}. For \ac{D-TDD}, the UL/DL direction is set in each cell based on the randomly arriving traffic with the mentioned distribution.

\begin{figure}
	\centering
	\includegraphics[width=0.75\textwidth]{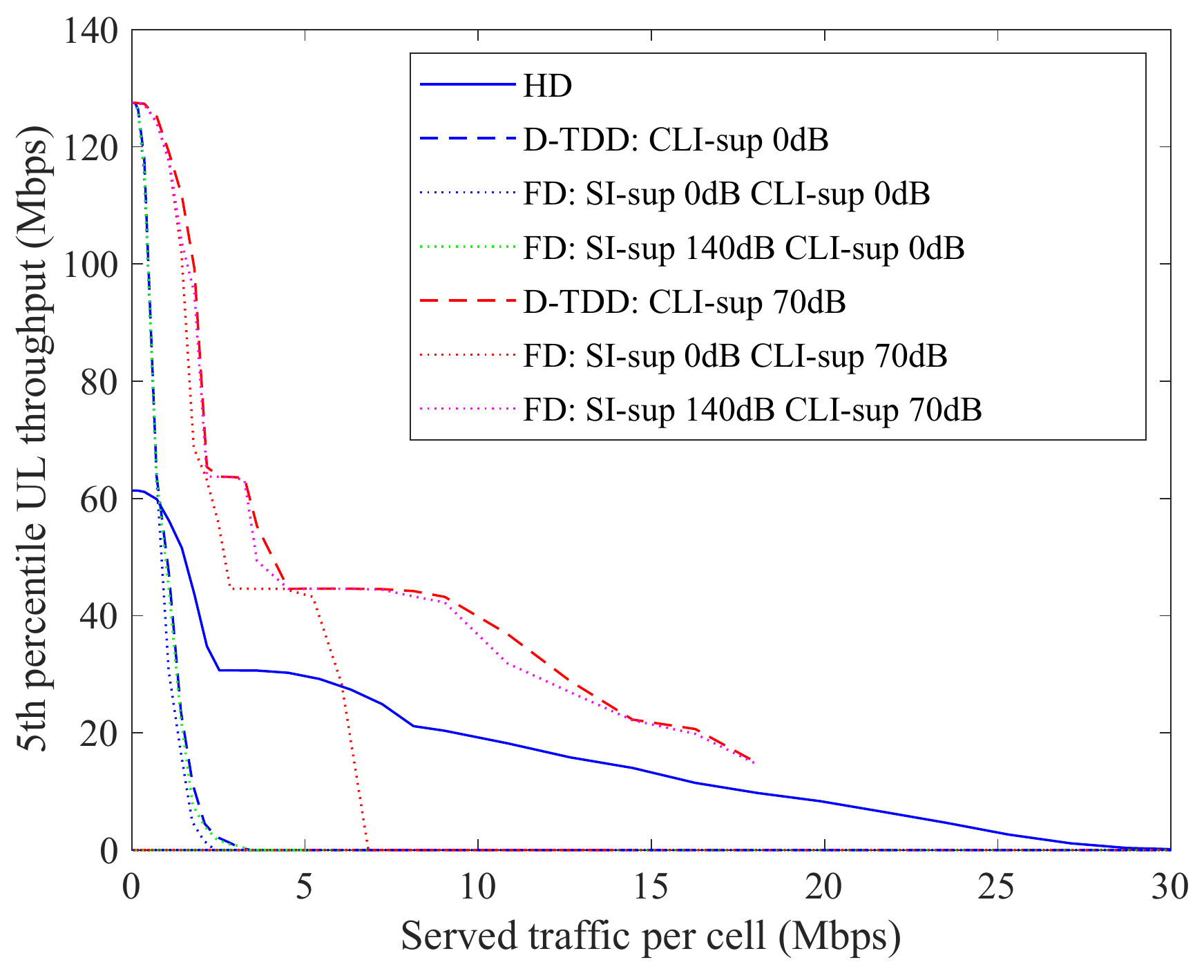}
	\caption{\ac{UL} throughput performance with \ac{FD} and \ac{D-TDD} with and without artificially suppressing \ac{SI} and \ac{CLI}, compared to \ac{HD}.}\label{fig:UL_throughput}
\end{figure}

The performance of \ac{HD} (static-\ac{TDD}/\ac{FDD}) network operating over paired $20 + 20$ MHz spectrum for \ac{UL} + \ac{DL} is contrasted to the performance of \ac{D-TDD} and \ac{FD} with no suppression of \ac{CLI} and \ac{SI}.
The unsuppressed \ac{SI} is assumed to occur with $0$ dB loss intra-cell, and intra-site \ac{CLI}, i.e., the \ac{UL}-to-\ac{DL} user interference, is assumed with $60$ dB loss, corresponding to TX and RX side lobe levels.
Figure~\ref{fig:UL_throughput} shows the UL throughput at the cell-edges (5\%-point)
as function of the increasing \ac{UL} traffic, which shows that interference degrades the performance already at lower-traffic loads, and is a more important factor than bandwidth.
When the \ac{CLI} is artificially suppressed by $70$ dB, the \ac{D-TDD} performance improves significantly, while the \ac{FD} performance improves only partly and requires an additional \ac{SI} suppression of $140$ dB for performance on par with \ac{D-TDD}.
\REVFR{Moreover, a \ac{SI} suppression of as much as \unit{140}{dB} without \ac{CLI} suppression does not provide performance gains.}
Therefore, mitigation of both \ac{CLI} and \ac{SI} is key to improve the performance of \ac{FD}.

\subsection{Performance under interference management}\label{sub:UMi_results}

Next, we characterize the system-level performance of \ac{FD} system featuring interference suppression in a favorable urban-macro scenario, where 7 BSs are arranged \unit{200}{m} apart in a hexagonal-grid. 
Each \ac{BS} is equipped with a $128$-antenna array directed towards the served \ac{UE} \REVFR{with $2$-antennas and $1$-stream}, corresponding to a large cylindrical array in practice.
The \ac{BS} antenna height is $10$ m and transmit power is $1$ W, while the carrier frequency is $3.5$ GHz and the 
system bandwidth is $40$ MHz. The frequency splitting between \ac{FD}, \ac{D-TDD}, and \ac{HD} is the same as in 
Section~\ref{sub:UMa_results}, and the path-loss and Rayleigh fading models follow the guidelines in~\cite{ITURM2135}. 
\REVFR{The \ac{SI} channel is modelled as Ricean fading with strong line-of-sight link and \ac{SI} cancellation of \unit{110}{dB}, where high cancellation has been shown experimentally for MIMO~\cite{Zhang2016,Everett2016}.}
Equal power allocation is considered in the \ac{DL} with a \ac{SINR} cap of \unit{30}{dB} while \ac{UL} power is as 
described in Section~\ref{sub:UMa_results}.

We assume traffic loads follow an M/M/1 queue modeling and can be divided as medium and low, by varying the probability that a user participates in the transmission. 
The medium- and low-traffic settings correspond to \ac{UL}-\ac{DL} cell utilization of $50\%$ and $10\%$, respectively. 
\REVFR{We assume a scheduling policy such that one \ac{UL} user and one \ac{DL} user share the time-frequency resource 
without any multi-user intra-cell interference.
The remaining inter- and intra-cell \ac{CLI} between users and the \acp{BS} are present due to the sharing of the same time-frequency resource in the same or across neighboring cells.}
\REVFR{Zero forcing (\acs{ZF}\acused{ZF})} precoder is used at the transmitter for \ac{FD}, \ac{D-TDD}, and \ac{HD}.
For \ac{D-TDD}, the \ac{UL} and \ac{DL} cells are assigned dynamically based on the user load.

\REVFR{We propose a \ac{CLI} interference management, termed \mbox{BSint}, based on the \ac{ZF} receiver utilizing the 
BS-to-BS link \ac{CSI}, which is obtained by means of pilots and information exchange via X2 interface.} 
Using this channel information, the receiver at the \ac{BS} points a null towards the direction of the \ac{BS}-to-\ac{BS} channel multiplied by the precoder of the interfering \ac{BS}. 
The spatial degrees-of-freedom are thereby used for interference suppression instead of scheduling multiple UEs. 
With this, the \ac{BS} can accurately cancel the \ac{BS}-to-\ac{BS} interference coming from at most $M$ \ac{DL} interfering \acp{BS}, where $M$ is the number of receive-antennas at the \ac{BS}, \REVFR{and without harming the received signal power as in~\cite{Choi2013}.}
Notice that \mbox{BSint} is a low-complexity solution that can be applied to either \ac{D-TDD} or \ac{FD}, and that it is much simpler than other \ac{CLI} management solutions analyzed in Section~\ref{sec:mimo_multi_cell}. 
Overall, we compare the performance of 7 algorithms (specifically, the $50$-th percentile of the curves): \ac{FD}, \ac{FD} with \mbox{BSint} on 4 and 6 \acp{BS}, \ac{D-TDD}, \ac{D-TDD} with \mbox{BSint} on 4 and 6 \acp{BS}, and \ac{HD}.

\begin{figure}
\centering
\subfloat[Comparison between \ac{FD} and \mbox{FD-4BSint} \label{fig:UL_interf_med_4BSint}] {\includegraphics[width=0.49\textwidth]
{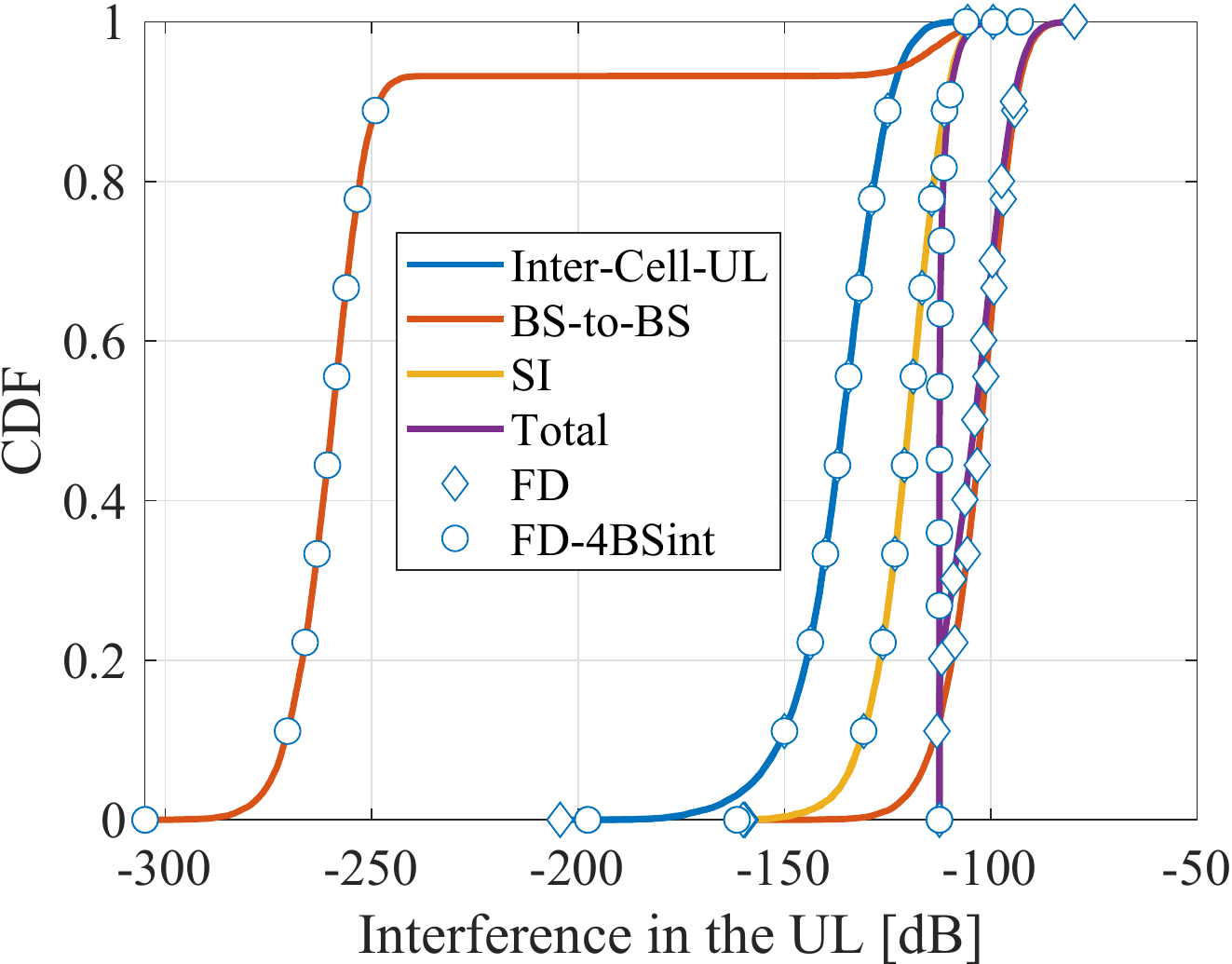}} \hfill
\subfloat[Comparison betwen \ac{FD} and \mbox{FD-6BSint}\label{fig:UL_interf_med_6BSint}] {\includegraphics[width=0.49\textwidth]
{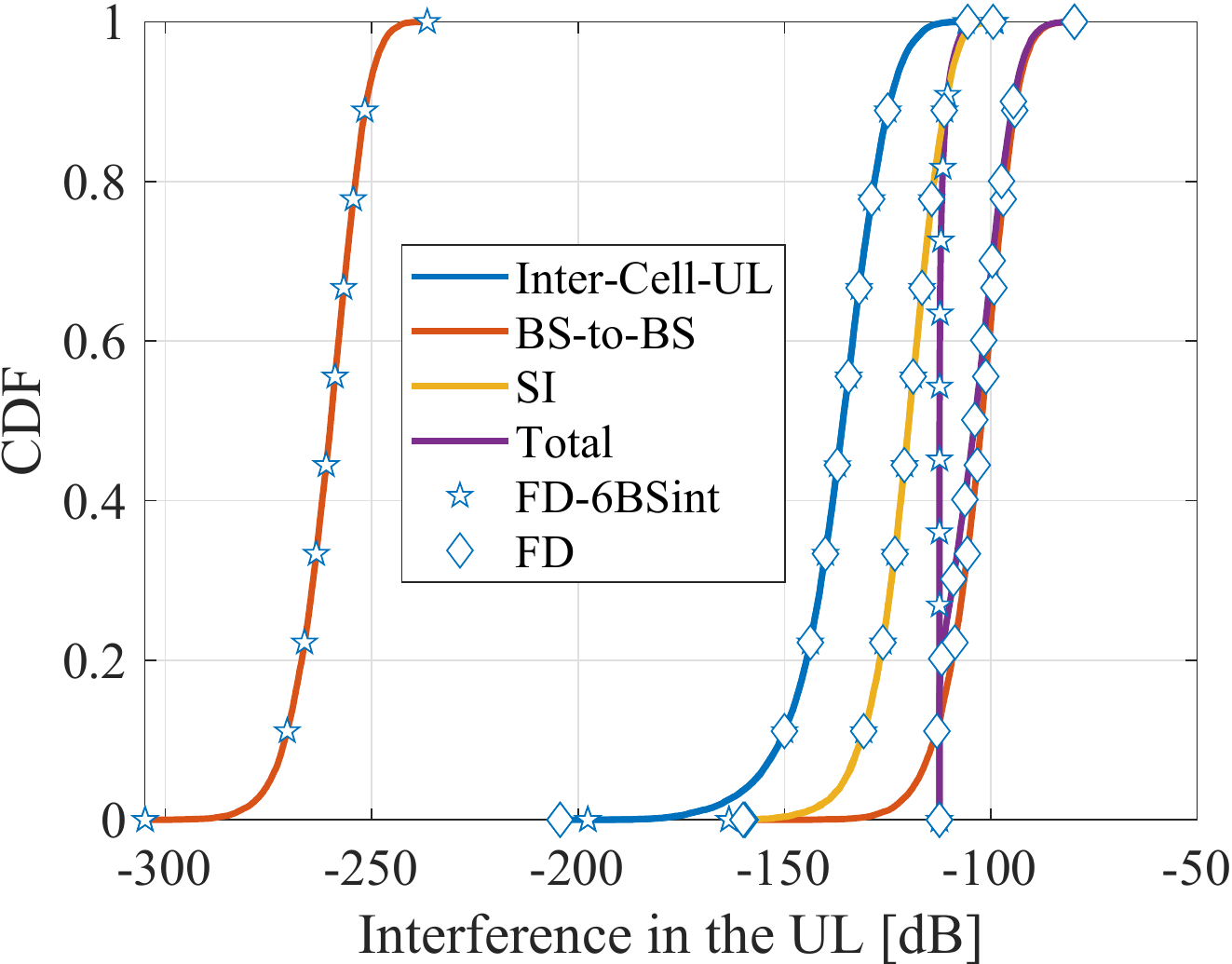}}
\caption{Comparison of the \ac{UL} interference sources between \ac{FD}, \mbox{FD-4BSint}, and \mbox{FD-6BSint} for medium-traffic. }
\label{fig:UL_interf_med}
\end{figure}

Figure~\ref{fig:UL_interf_med} compares the \ac{CDF} of \ac{UL} interference for scenarios including \ac{FD}, \mbox{FD-4BSint}, and \mbox{FD-6BSint}.
Specifically, Figure~\ref{fig:UL_interf_med_4BSint} evinces that the \ac{BS}-to-\ac{BS} interference, and not the \ac{SI}, is the highest interference source in the \ac{UL} of \ac{FD}.
With \mbox{FD-4BSint}, the \ac{BS}-to-\ac{BS} interference is completely suppressed in approximately \unit{86}{\%} of the cases.
Hence, there are at most $4$ interfering \acp{BS} in each resource in most cases.
For \mbox{FD-6BSint}, the \ac{BS}-to-\ac{BS} interference is fully cancelled, i.e., the \ac{CLI} curve is far in the left.
The results show that for a medium-traffic, the \mbox{FD-4BSint} solution is sufficient to cancel the vast majority of the \ac{UL} \ac{CLI}, and that after its cancellation, the noise and \ac{SI} become the strongest interference sources.
In the low-traffic scenario, \mbox{FD-4BSint} is expected to have a performance much closer to \mbox{FD-6BSint} than in the medium-traffic due to fewer interfering sources, which is not presented due to lack of space.

\begin{figure*}
\centering
\subfloat[System Sum Throughput\label{fig:Sum_throughput_medium}] {\includegraphics[width=0.33\textwidth]
{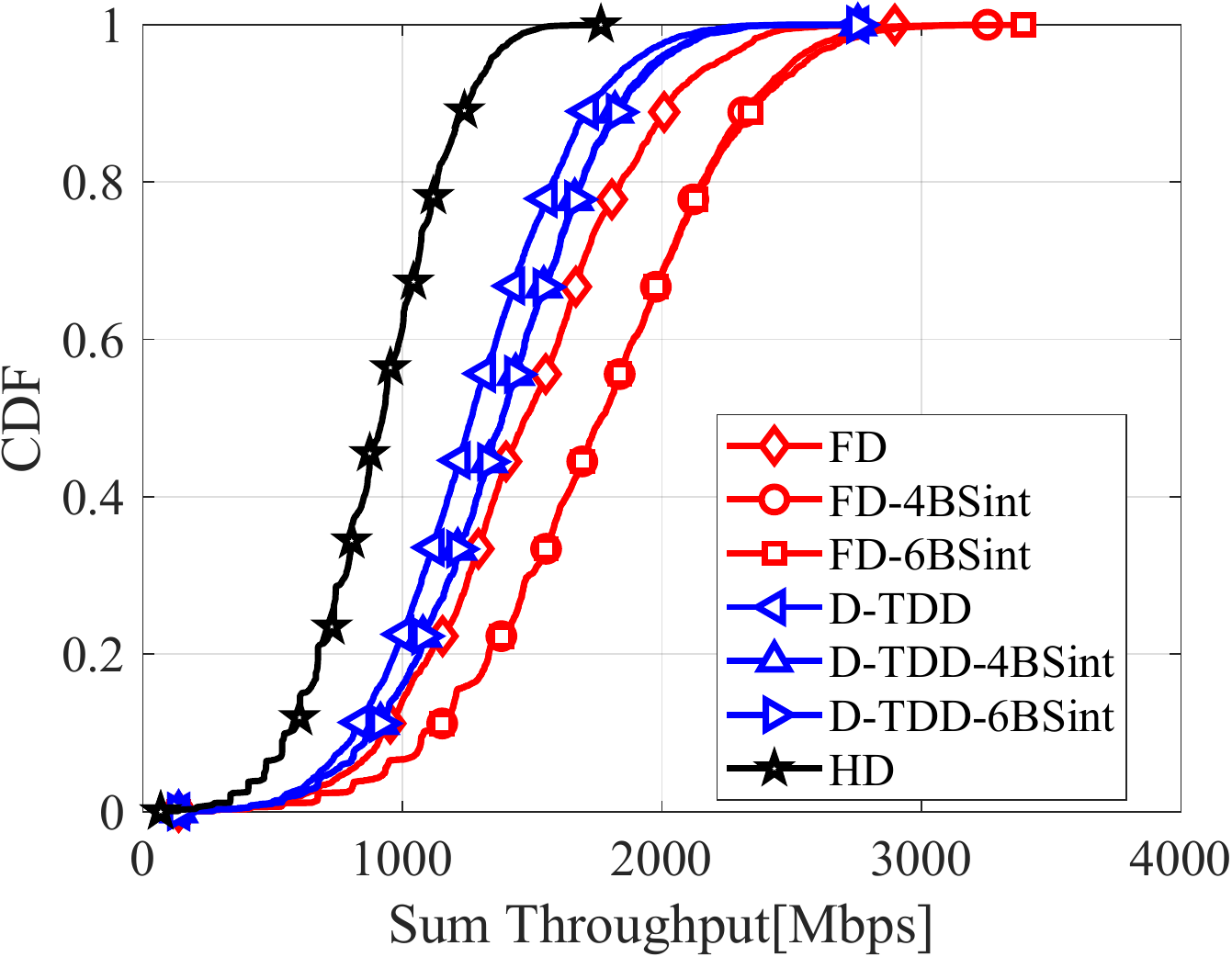}} \hfill
\subfloat[Throughput \ac{UL}\label{fig:UL_throughput_medium}] {\includegraphics[width=0.33\textwidth]
{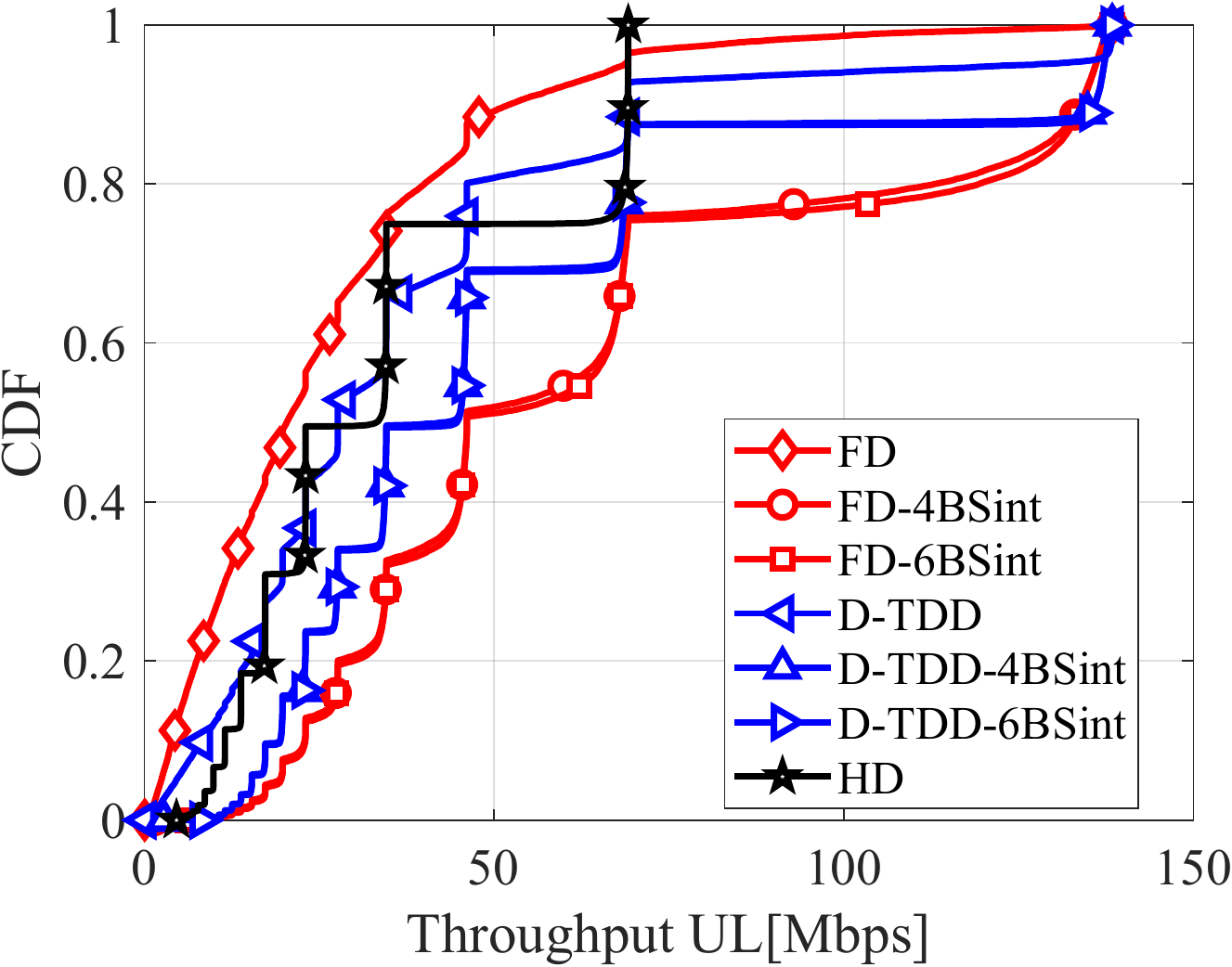}} \hfill
\subfloat[Throughput \ac{DL}\label{fig:DL_throughput_medium}] {\includegraphics[width=0.33\textwidth]
{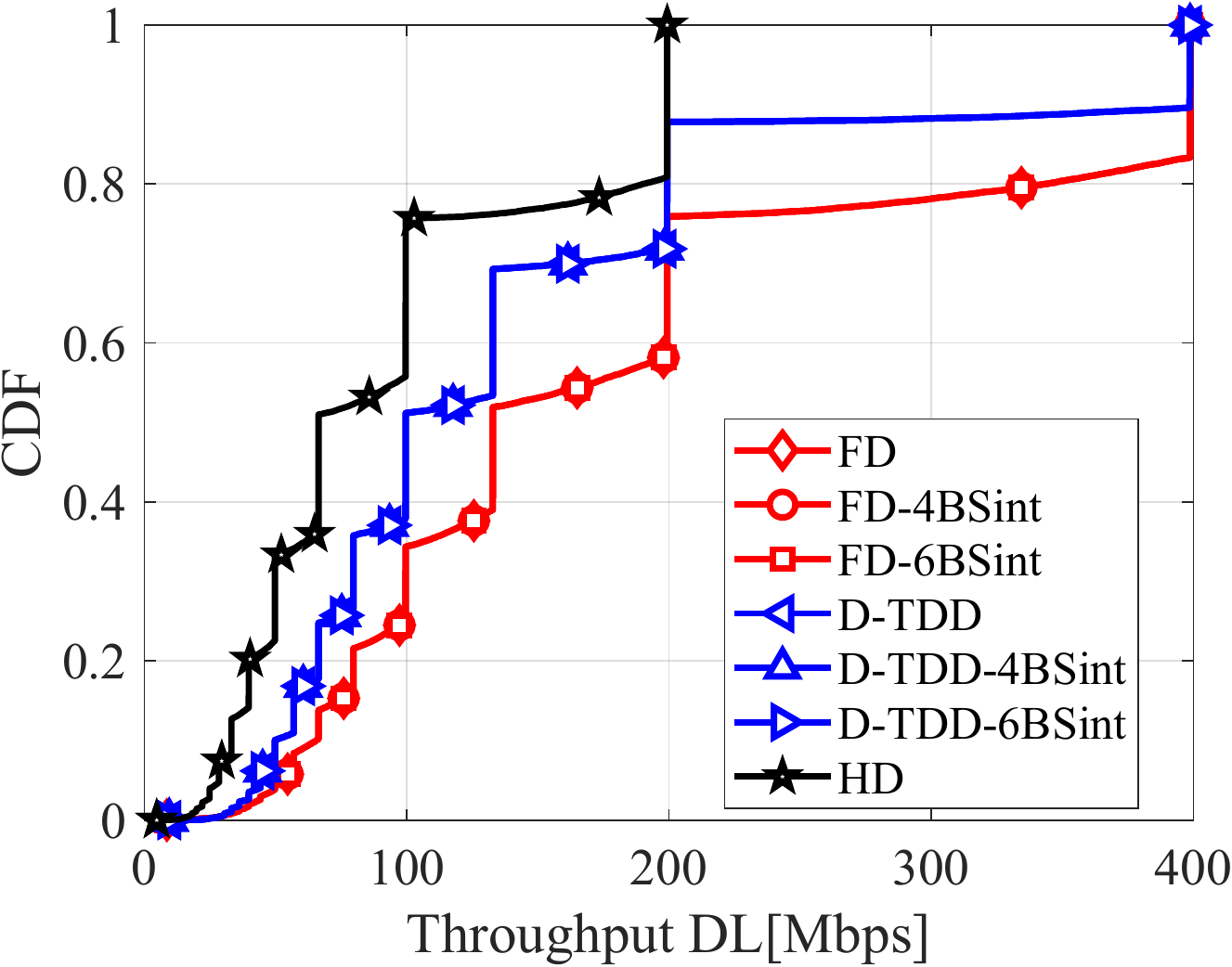}}
\caption{Throughput for the system's sum throughput, and individual \ac{UL} and \ac{DL} users in a medium-traffic scenario.}
\label{fig:Users_throughput_medium}
\end{figure*}
In Figure~\ref{fig:Users_throughput_medium}, we show the \ac{CDF} of the system, and the individual throughput in a medium-traffic scenario for both UL and DL.
\ac{FD} transmissions outperform \ac{HD} and the \ac{D-TDD} solutions (cf. Figure~\ref{fig:Sum_throughput_medium}).
The relative throughput gain between \ac{FD} and \ac{HD} is \unit{60}{\%}; whereas the relative gain to \ac{D-TDD} is \unit{17}{\%}.
Using the \mbox{FD-4BSint} and \mbox{FD-6BSint} solutions, \ac{FD} provides further gains over\ac{HD} and \ac{D-TDD}.
The relative throughput gains are \unit{91}{\%} and \unit{39}{\%}, respectively, and there is not much difference between \mbox{FD-4BSint} and \mbox{FD-6BSint}.
This behavior happens because the \acp{BS} interference is almost cancelled when \mbox{FD-4BSint} is used (cf. Figure~\ref{fig:UL_interf_med}).

For the \ac{UL} users in Figure~\ref{fig:UL_throughput_medium}, \ac{FD} is outperformed by \ac{HD} and all \ac{D-TDD} solutions. 
For instance, the relative difference between \ac{FD} and \ac{HD} is \unit{38}{\%} while the difference to \ac{D-TDD} is \unit{27}{\%}.
Nevertheless, \mbox{FD-4BSint} and \mbox{FD-6BSint} outperform \ac{HD} and \ac{D-TDD}, in \unit{39}{\%} and \unit{67}{\%}, respectively; whereas their performance gains are \unit{3}{\%} to the \ac{D-TDD} solutions with \ac{CLI} cancellation.
With proper \ac{BS}-to-\ac{BS} interference suppression, \ac{FD} transmissions provide higher throughput for the \ac{UL} users than \ac{HD} and all \ac{D-TDD} solutions.
For the \ac{DL} users in Figure~\ref{fig:DL_throughput_medium}, the curves have a staircase shape because of the \ac{SINR} cap and the frequency splitting.
The relative throughput gains of \ac{FD} transmission compared to \ac{HD} and \ac{D-TDD} are \unit{73}{\%} and \unit{16}{\%}, respectively.
The gains with \ac{FD} are much higher in the \ac{DL}, which shows that the \ac{UL}-to-\ac{DL} \ac{CLI} is not a limiting factor on the performance of \ac{FD}.
\REVFR{The reasons for this behavior include: the \acp{UE} have directional transmissions, transmitted power designed 
to target \unit{10}{dB} \ac{SNR} with lower maximum transmit power than the \ac{BS}, and lower elevation, thus 
increasing the pathloss among \acp{UE}; and, importantly, the \acp{UE} have a lower probability of a certain nearby 
\ac{UE} being active in a certain resource block.}
Since the interference suppression is focused only on the \ac{BS}-to-\ac{BS} interference at the \ac{UL}, \mbox{FD-4BSint} and \mbox{FD-6BSint} perform as the \ac{FD}.
Hence, in the medium-traffic \ac{FD} provides high sum throughput gains compared to \ac{HD} and \ac{D-TDD}, and that \ac{FD} improves the throughput of both \ac{UL} and \ac{DL}.
With proper \ac{CLI} management, the gains are even higher and approach the desired doubling of the sum throughput.

\begin{figure*}
\centering
\subfloat[System Sum Throughput\label{fig:Sum_throughput_low}] {\includegraphics[width=0.33\textwidth]
{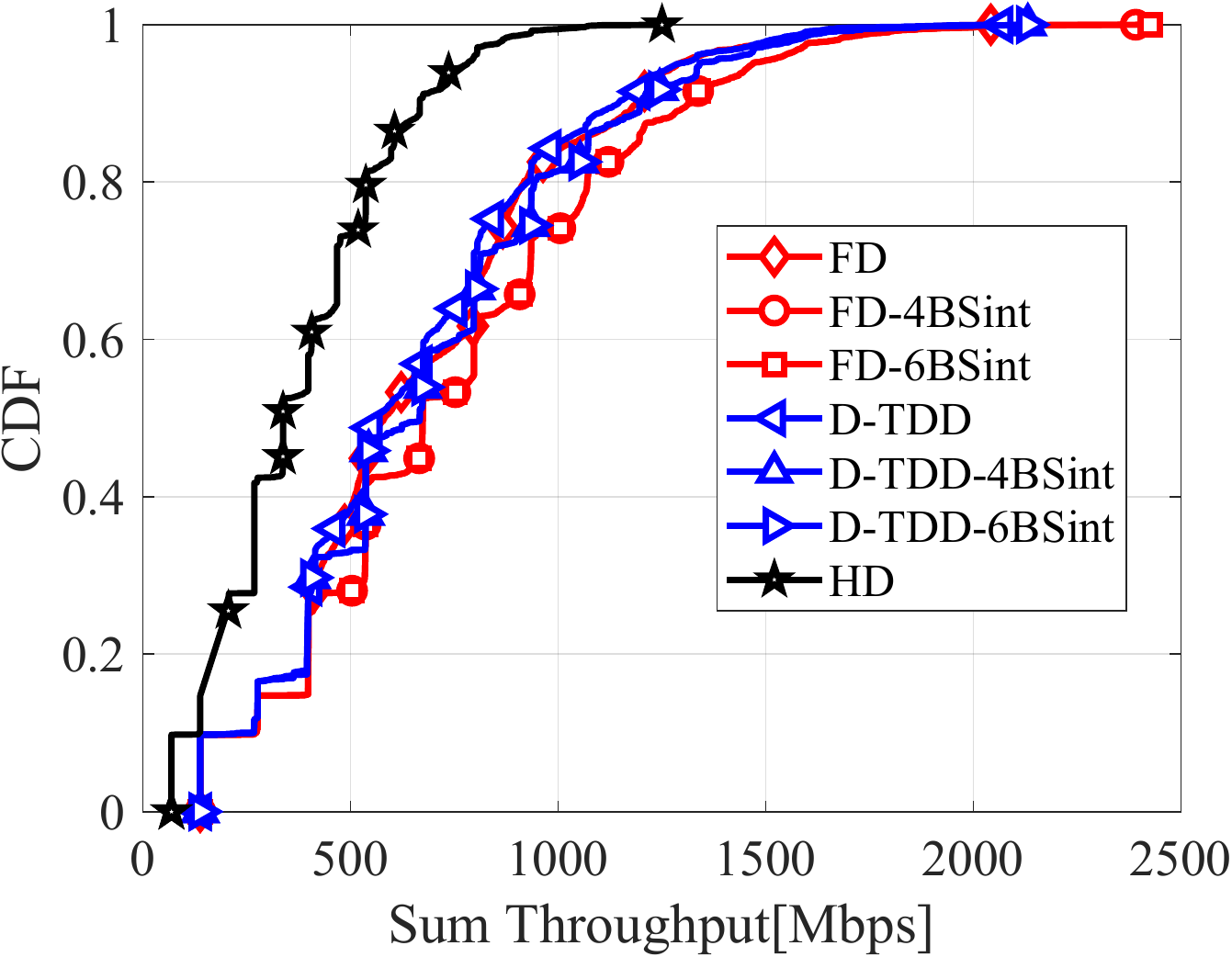}} \hfill
\subfloat[Throughput \ac{UL}\label{fig:UL_throughput_low}] {\includegraphics[width=0.33\textwidth]
{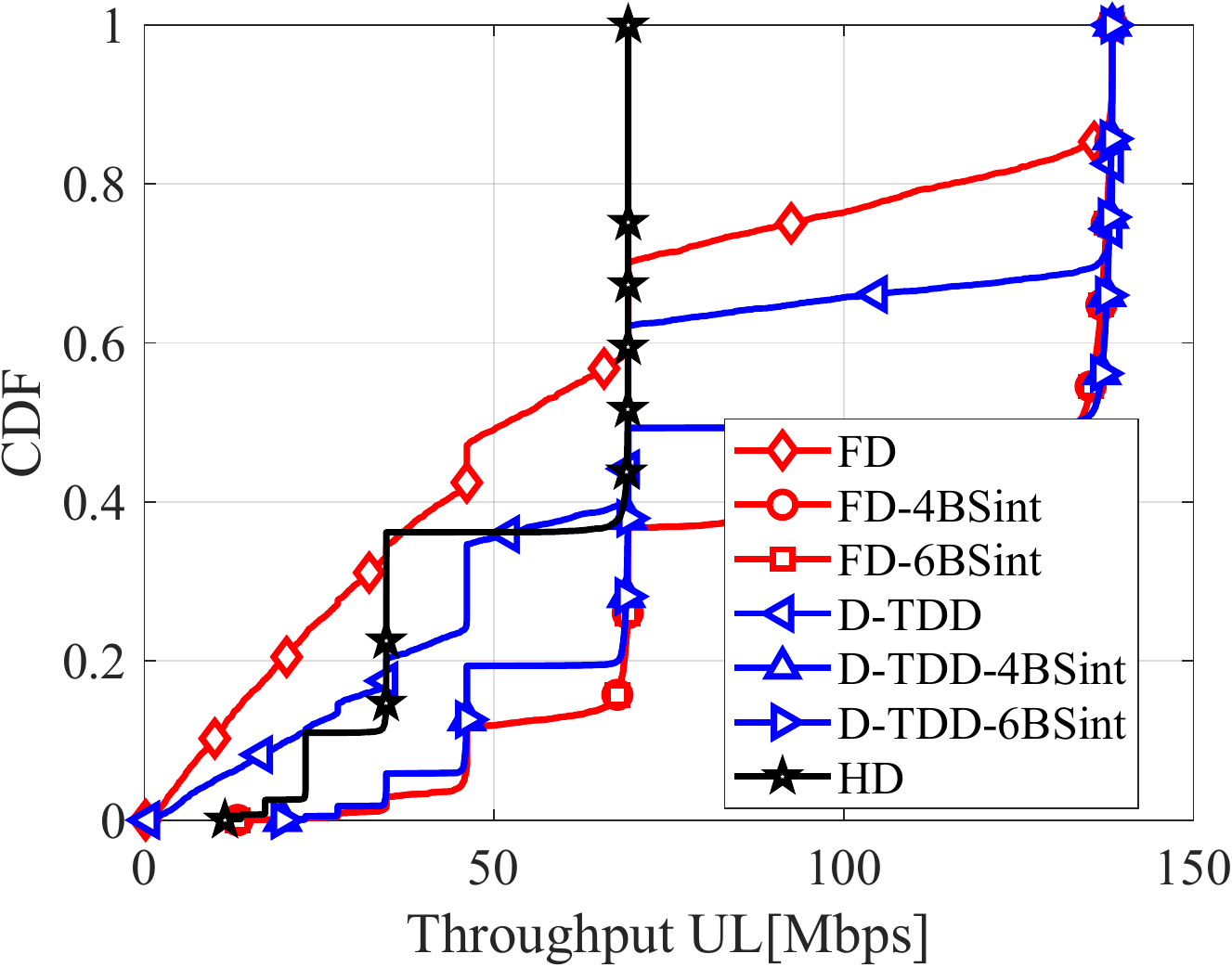}} \hfill
\subfloat[Throughput \ac{DL}\label{fig:DL_throughput_low}] {\includegraphics[width=0.33\textwidth]
{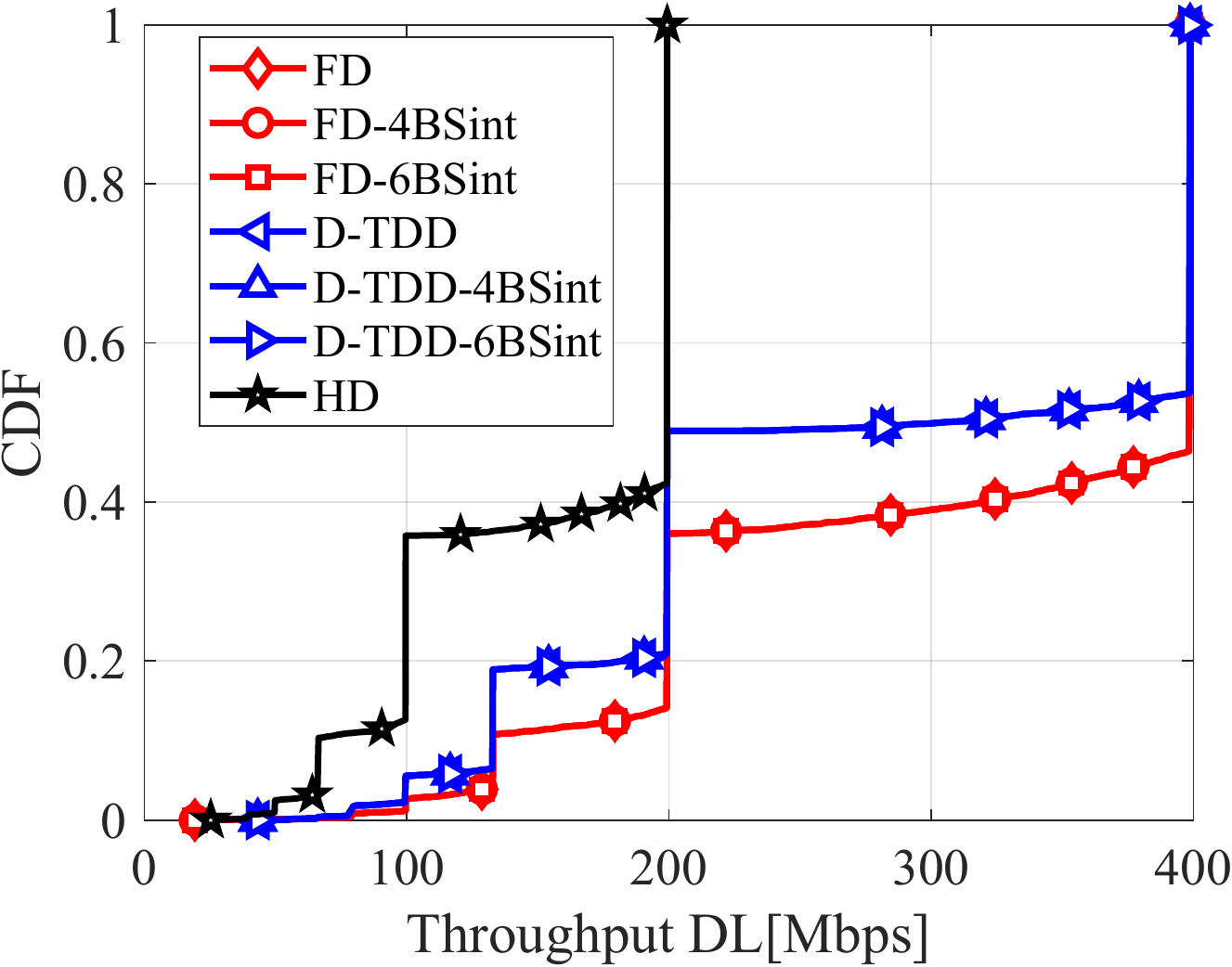}}
\caption{Throughput for the system's sum throughput, and individual \ac{UL} and \ac{DL} users in a low-traffic scenario.}
\label{fig:Users_throughput_low}
\end{figure*}
In Figure~\ref{fig:Users_throughput_low}, we show the \ac{CDF} of the system, and the individual \ac{UL} and \ac{DL} users throughput in a low-traffic scenario.
Similar to the medium-traffic, in Figure~\ref{fig:Sum_throughput_low} \ac{FD} solution outperform \ac{HD} and \ac{D-TDD} solutions.
However, the throughput gains compared to the \ac{D-TDD} solutions is much lower now, \unit{4}{\%}; whereas the gains to \ac{HD} are high, \unit{72}{\%}.
The reason for high gains compared to \ac{HD} is the resource utilization in the low-traffic.
Most resources are occupied by a single \ac{UL} or \ac{DL} user, which uses the whole bandwidth of \unit{40}{MHz} instead of the bandwidth of \unit{20}{MHz}.
Due to the same reason above, the performance gains compared to \ac{D-TDD} is much lower than in the medium-traffic, and both \mbox{FD-4BSint} and \mbox{FD-6BSint} have the same performance in terms of sum throughput.
Nevertheless, the relative gain is \unit{20}{\%} between \ac{FD}, \mbox{FD-4BSint} and \mbox{FD-6BSint}.
Moreover, \mbox{FD-4BSint} and \mbox{FD-6BSint} have a relative gain of approximately \unit{99}{\%} when compared to \ac{HD}, which shows that \ac{CLI} management yields almost doubling of the sum throughput.
\ac{FD} transmissions benefit from the low-traffic scenario due to low resource usage between \ac{UL} and \ac{DL} users, and the \ac{CLI} management further improves these benefits.

For the \ac{UL} users in Figure~\ref{fig:UL_throughput_low}, \ac{FD} is still outperformed by \ac{HD} and all \ac{D-TDD} solutions.
As in Figure~\ref{fig:UL_throughput_medium}, the solutions \mbox{FD-4BSint} and \mbox{FD-6BSint} perform much better than \ac{FD}.
Their relative performance gains compared to \ac{HD} and \ac{D-TDD} are \unit{93}{\%}.
When comparing \mbox{FD-4BSint} and \mbox{FD-6BS} with \mbox{D-TDD-4BSint} and \mbox{D-TDD-6BSint}, the relative gains are much smaller and close to \unit{1}{\%}.
This behavior shows that in the low-traffic and enough \ac{CLI} cancellation, the \ac{UL} performance of \ac{FD} and \ac{D-TDD} are almost the same.
Hence, the \ac{BS}-to-\ac{BS} interference suppression provides higher gains in the low-traffic than in the medium-traffic.
For the \ac{DL} users in Figure~\ref{fig:DL_throughput_low}, the performance is similar to the medium-traffic in Figure~\ref{fig:DL_throughput_medium}.

\REVFR{We generated results for lower levels of \ac{SI} cancellation, such as \unit{90}{dB}, where we noticed that the 
\ac{BS}-to-\ac{BS} interference is still crucial together with the \ac{SI}. Due to space limitations, we do not include 
these results.}
\REVFR{Neither the system nor the \ac{DL} throughput are impacted, only the \ac{UL} throughput that is slightly outperformed by \ac{D-TDD} with \ac{BS}-to-\ac{BS} interference cancellation. 
Hence, cancellations $\leq 90$ dB limit only the UL performance.}

Overall, the \ac{BS}-to-\ac{BS} interference is the highest source of interference in a multi-cell scenario 
\REVFR{provided the \ac{SI} is sufficiently cancelled}. 
With proper \ac{BS}-to-\ac{BS} interference, \ac{FD} brings throughput gains for both \ac{UL} and \ac{DL} users in different traffic scenarios and can almost double the throughput for the system and individual users. 
\section{Conclusions, Lessons Learned and Perspective}\label{sec:concl}

Recently, \ac{D-TDD} and \ac{FD} have appeared as technologies that can push the reuse from 1 to $1/2$.
Despite the similarities between \ac{FD} and \ac{D-TDD}, 
the application of \ac{FD} transmissions in cellular systems has not advanced yet due to the harsher interference situation.
In this paper, we argued that the key limitation is not the \ac{SI}, but the \ac{CLI} on the \ac{UL}, i.e., the \ac{BS}-to-\ac{BS} interference, which we show validated by means of simulations.
To support a smooth implementation of \ac{FD} transmissions in current networks, we discussed the relevant studies and signaling measurements already standardized for \ac{D-TDD}.
Using these ideas, we proposed a low-complexity receiver that mitigates the CLI  at the receiving BS by exploiting the 
BS-to-BS link \ac{CSI}. 
We showed that \ac{FD} under \ac{CLI} management greatly outperforms \ac{HD} and \ac{D-TDD} for medium- and low-traffic 
scenarios in terms of system and individual throughput for \ac{UL} and \ac{DL} users.

From our results, we can summarize the lessons learned as follows:
\begin{enumerate}
    \item With the high \ac{SI} cancellation achievable in current \ac{MIMO} systems, the \ac{BS}-to-\ac{BS} interference, and not the \ac{SI} or \ac{UL}-to-\ac{DL} interference, is the limiting factor in \ac{FD} cellular networks;
    \item Medium- and low-traffic scenarios benefit \ac{FD} transmissions due to the efficient spectrum reuse across cells;
    \item \ac{D-TDD} also gives efficiency gains, especially in low-traffic scenarios;
    \item \ac{CLI} management is essential for providing high UL throughput gains;
    \item Low-complexity receivers suitable to both \ac{D-TDD} and \ac{FD} are possible and may help to realize FD cellular networks.
\end{enumerate}

To achieve the \ac{CLI} management discussed in this paper and realize efficient \ac{D-TDD} and \ac{FD} in a 5G multi-cell system, measurements between \ac{BS}s would be required to estimate \ac{CSI}. 
Recent work on \ac{CLI} \cite{3gpp.38.828} has not resulted in specification of the required enabler techniques in terms of coordination or measurements. 
Given that \ac{BS}s are stationary and interference is co-channel, both direct measurements on probes from the victim BS in the source \ac{BS} and vice-versa 
and exchange of results, appears feasible. 
The high-level of suppression calls for highly accurate \ac{ZF} and frequent estimates, which is possible with slight increase of overhead and computing costs. 

A different approach is to use the antenna correlation at the \acp{BS} to accurately estimate the strongest direction of the channel between each source-\ac{BS} and the victim-\ac{BS}.
Using this idea, it is not necessary to account for the beamforming of each \ac{DL} user in the source-\ac{BS}.
\REVFR{This approach is similar to~\cite{Choi2013}, but more specific due to use of the directions of interfering source-\acp{BS}, and without harming the received power.}
In situations that are complicated to obtain an accurate \ac{CSI}, 
a possibility is to estimate the channel from other source-\acp{BS} and form an average over their used beams towards the victim-\ac{BS}.

Interesting topics for future research comprise the importance of the \ac{CSI}, and the necessary accuracy in suppressing \ac{CLI} to realize efficient \ac{D-TDD} and \ac{FD} in a 5G multi-cell system. 
When this is achieved, along with accurate \ac{SI} suppression, \ac{FD} can become a next evolutionary step in spectrum reuse.
Another interesting topics for future study are \REVFR{the suitability of cell-free massive \ac{MIMO} and its central processing for \ac{CLI} suppression,} to see if \ac{FD} is a better use of the antenna degrees of freedom than \ac{MU-MIMO}, and if both techniques can be combined.
\bibliographystyle{IEEEtran}
\bibliography{FDref}

\end{document}